\documentclass[twocolumn]{autart}    

\usepackage{graphicx}          
\usepackage[english]{babel}
\usepackage{url} 

\usepackage{amsfonts, amsmath, mathtools}
\usepackage{txfonts} 
\usepackage{subfig} 
\usepackage{cite}  
\usepackage{enumerate}

\newif\iftikz
\tikztrue 

\iftikz
\usepackage{tikz}
\usetikzlibrary{arrows,calc,decorations.pathmorphing,decorations.pathreplacing, decorations.markings,backgrounds,positioning,fit, shapes.geometric, patterns}
\fi

\usepackage{enumitem} 
\usepackage{color} 

 \usepackage{hyperref}  

\newtheorem{con}[thm]{Conjecture}
\newtheorem{ass}[thm]{Assumption} 

\usepackage{parskip}
\newcommand\ie{i.\,e.}

\newcommand\hp{h_\mathrm p}
\newcommand\hd{h_\Delta}
\newcommand\BB{{\mathcal{B}}} 
\newcommand\BBT{{\mathcal{B}}^\mathrm T}

\newcommand\BBdt{\tilde{\mathcal{B}}_\Delta}

\newcommand\BBpt{\tilde{\mathcal{B}}_\mathrm p}

\newcommand{\Babs}{\langle \BB \rangle}
\newcommand{\hinfnorm}{\mathcal{H}_\infty}
\newcommand{\smin}{\sigma_{\min}}
\newcommand{\smax}{\sigma_{\max}}
\newcommand{\lmin}{\lambda_{\min}}
\newcommand{\lmax}{\lambda_{\max}}

\newcommand{\T}{\mathrm{T}}
\newcommand{\Tf}{T_{\mathrm f }} 

\newcommand{\lp}{{L_p}}

\newcommand{\vt}{\tilde v}

\renewcommand\d{\mathrm{d}}

\newcommand{\dreference}{{\delta_{\text{ref}}}}

\begin{document}
\iftikz

\tikzstyle{genGraphNode} = [minimum size=3mm, thick, node
distance=5mm] 
\tikzset{
    >=latex, bend angle=10 }
\tikzstyle{normalNode} = [genGraphNode, circle,draw=black,fill=black!10,thick]
\tikzstyle{controllingNode} = [genGraphNode,
circle,draw=black!20,fill=black!60,very thick, text=white, font=\bfseries] 
\tikzstyle{observingNode} =[genGraphNode, circle,draw=black!50,fill=white,thick]
\tikzstyle{placeholder}=[genGraphNode, circle]

\tikzstyle{block} = [draw, fill=white!20, rectangle, 
    minimum height=1em, minimum width=1.5em] 
\tikzstyle{gain} = [draw,regular polygon,
        regular polygon sides=3, minimum height=2em, minimum
width=1em, inner sep=0pt] 
\tikzstyle{gainLeft} = [gain, shape border rotate=-90]
\tikzstyle{gainRight} = [gain, shape border rotate=90]
\tikzstyle{gainDown} = [gain, shape border rotate=180]
\tikzstyle{gainUp} = [gain, shape border rotate=90]
\tikzstyle{sum} = [draw, fill=white!20, circle,inner sep=1pt]
\tikzstyle{input} = [coordinate]
\tikzstyle{output} = [coordinate]
\tikzstyle{pinstyle} = [pin edge={to-,thin,black}]
\tikzstyle{junction} = [draw, circle, minimum height=0.02em, fill=black, inner sep=0pt]
\fi
\begin{frontmatter}

\title{Disturbance scaling in bidirectional vehicle platoons with different asymmetry in position and velocity coupling}
 
\author[P]{Ivo Herman}, \ead{ivo.herman@fel.cvut.cz}
\author[UU]{Steffi Knorn} \ead{steffi.knorn@signal.uu.se} and
\author[UU]{Anders Ahl\'en} \ead{anders.ahlen@signal.uu.se} 
\address[P]{Department of Control Engineering, Czech Technical University in Prague, Prague, Czech Republic}
\address[UU]{Signals and Systems, Uppsala University, Sweden} 
 
\begin{abstract}
  This paper considers a string of vehicles where the local control law uses the states of the vehicle's immediate predecessor and follower. The coupling towards the preceding vehicle can be chosen different to the coupling towards the following vehicle, which is often referred to as an asymmetric bidirectional string. Further, the asymmetry for the velocity coupling can be chosen differently to the asymmetry in the position coupling. 
  It is investigated how the effect of disturbance on the control errors in the string depends on the string length. It is shown, that in case of symmetric position coupling and asymmetric velocity coupling, linear scaling can be achieved. For symmetric interaction in both states, i.e., in symmetric bidirectional strings, the errors scale quadratically in the number of vehicles. When the coupling in position is asymmetric, exponential scaling may occur or the system might even become unstable. The paper thus gives a comprehensive overview of the achievable performance in linear, asymmetric, bidirectional platoons. The results reveal that symmetry in the position coupling and asymmetry in velocity coupling qualitatively improves the performance of the string. Extensive numerical results illustrate the theoretical findings.
\end{abstract}


\end{frontmatter}

%

\section{Introduction}\label{sec_intro}

Vehicle platoons form an important part of future intelligent transportation systems, because such systems are anticipated to increase both the safety and capacity of highways. In its simplest form, a platoon, consisting of $N$ cooperatively-acting, automatically controlled vehicles travels in a longitudinal line with tight spacing between the vehicles. 

An important safety and performance measure in this area is how the response of the platoon to disturbances scales with respect to the number of vehicles $N$. When the local errors are bounded independently of $N$, the string is called ``string stable''. Generally speaking, a platoon is string stable if disturbances, which are propagating through the string, do not grow with the number of vehicles or the position within the string. See \cite{Ploeg2014} for various definitions of string stability.

The literature often distinguishes between ``unidirectional'' strings, where each vehicle only considers information of a group of direct predecessors, and ``bidirectional'', where information from following vehicles is also used. It is well known, that a strict form of string stability in linear vehicle strings with double integrators in the open loop, local information only and tight spacing, can neither be achieved in unidirectional nor in bidirectional strings \cite{Seiler2004a, Barooah2005}. {This definition of string stability requires the $L_2$ norm of the local error vector to be bounded for any $L_2$ bounded disturbance vector.}
{String stability according to this definition in unidirectional strings can for instance be achieved using a time headway spacing policy, \cite{Middleton2010} or by allowing inter-vehicle communication, \cite{Alam2015}. However, the time-headway policy leads to undesirable large steady-state inter vehicle distances and wireless communication between the vehicles can potentially be disturbed by an intruder.}  

Bidirectional strings seem to offer advantages since also backward distance errors are used to control the vehicle's motion. A bidirectional system, where the control input due to the forward distance error is weighed as high as the backward distance error, is referred to as a ``symmetric'' string. For example, a weaker form of string stability, can be achieved \cite{Knorn2014}, {where the $L_\infty$ norm of the local error vector is guaranteed to be bounded for any disturbances in $L_2$. The results were extended to analyse the effects of measurement errors in \cite{Knorn2015}.} Further, using similar tools allowed the analysis of a system with a more general graph topology for the inter vehicle connections \cite{Knorn2016}. In contrast, by weighing the forward error higher, that is, allowing asymmetric controller gains, some important benefits can be obtained. A uniform bound on the eigenvalues of the formation can be achieved \cite{Hao2012b}, which guarantees much faster transients than what are achievable with symmetric control.

However, the transients are not solely determined by the eigenvalues. The price to pay for the better convergence rate is that the $\hinfnorm$ norm of the transfer functions in the platoon scales exponentially in $N$ for asymmetric strings (compared to linear scaling for symmetric strings \cite{Veerman2007}). {This extremely bad scaling was first shown for a double-integrator model in \cite{Tangerman2012} and later generalized to an arbitrary agent model in \cite{Herman2013b} and to an arbitrary transfer function in \cite{Herman2014b}.}

Many works in the area assume that the degree of asymmetry in the position and velocity coupling is identical, see for instance \cite{Barooah2009a}. However, the performance of the string can be improved by assuming symmetric position coupling and asymmetric velocity coupling. { Good stability margin properties were shown in \cite{Hao2010}.} A good scaling of system norms with such an approach was numerically shown in \cite{Hao2012c} and the properties of the platoon's response to a step change in leader's velocity were derived in \cite{Cantos2014a}. Using these properties, the parameters of the controller as well as the coefficient of asymmetry can be optimised to minimise the transient time \cite{Herman2016}. In both papers \cite{Cantos2014a, Herman2016} it was shown that symmetry in position is necessary for a good scaling. In addition, it was proved in \cite{Martinec2015b} that symmetry in position is a necessary condition for ``local'' string stability. Note that, when different asymmetries in velocity and position  coupling are used, none of the convenient approaches presented in the literature for a distributed system analysis (e.g.,\cite{Fax2004a}) and synthesis (e.g.,\cite{Li2011}) can be used since the Laplacians for position and velocity coupling are not simultaneously diagonalisable.

\subsection{Problem formulation}
{This paper considers a heterogeneous, asymmetric, bidirectional string of $N$ vehicles with constant masses $m_i$, positions $x_i$, velocities $v_i$ and momenta $p_i$ for all $i\in\{1,2,\dots,N\}$. The vehicles are modelled as double integrators  such that
\begin{equation}\label{eq_xddot}
	m_i \ddot x_i = F_i +d_i, 
	\end{equation}
where $d_i$ is the disturbance acting on vehicle $i$. The linear control force $F_i$ has the form
\begin{align}\label{eq_Fi}
	\nonumber F_i =& (1+\hp)r_i(v_{i-1}-v_{i}) - (1-\hp)r_{i+1}(v_{i}-v_{i+1}) \\
	&+ (1+\hd)a_i\Delta_i -(1-\hd)a_{i+1}\Delta_{i+1},
\end{align}
where $\Delta_i := x_{i-1}-x_i-\dreference_{i-1, i}$ is the local position error between vehicle $i$ and its predecessor $i-1$, aimed to be kept at the fixed distance $\dreference_{i, j}$; constants $r_i>0$ and $a_i>0$  are the velocity and position coupling parameters and $\hp$ and $\hd$ are the asymmetry coefficients for the velocity and position coupling, respectively. The index $i=0$ refers to a virtual reference vehicle with position $x_0$, velocity $v_0$ and momentum $p_0$ and the last vehicle only considers the forward error as it has no follower.} 

{This paper investigates the disturbance scaling for such asymmetric bidirectional platoons. Collecting all states of the platoon in the vector $\chi(t)$ and all disturbances in the vector $d(t)$, disturbance scaling refers to how the scaling factor $\xi$ in $|\chi(t)|_\infty\leq |\chi(0)| +\xi \|d(\cdot)\|_2$ scales with the string length $N$, where $\|d(\cdot)\|_2$ denotes the $L_2$ norm of the disturbances. Specifically, the paper focuses on how $\xi$ scales with $N$ for different choices of the asymmetry parameters $\hp$ and $\hd$.}

\subsection{Contributions}
\begin{figure}
\centering
\iftikz
\begin{tikzpicture}[scale=0.7, >=latex]
\fill[white, pattern=north west lines, pattern color=blue!30] (0, 0) -- (3, 3) |- ( 6, 3) -- (6, 0) -- cycle;
	\fill[white, pattern=north east lines, pattern color=blue!5] (3, 3) -- (6,6) |- ( 6, 3)  -- cycle;
	\fill[white, pattern=north west lines, pattern color=red!20] (0, 0) -- (6,6) |- ( 0, 6)  -- cycle;

	\draw[->] (0,0) -- (6,0) node[anchor=north] {$\hp$};
	\draw[->] (0,0) -- (0,6) node[anchor=east] {$\hd$};
	\draw[-|] (2.9,0) -- (3,0) node[anchor=south, below=0.1cm]
	{$1$};
	\draw [-, line width=2mm, dash pattern=on 3pt off 3pt, color=black!30!green, dashed] (0,0) --  (5.8, 0) ;
	\draw[-|] (0, 2.9) -- (0, 3) node[anchor=west, left=0.1cm]
	{$1$};
	\draw[-] (0,0) -- (6, 6);
	\draw[-|] (0,0) -- (3,3) node[above=0.1cm] {PF};
	\draw (3,3) -- (6,3);
	\draw [red, ultra thick] (0.2, 0.2) -- (-0.2, -0.2);
	\draw [red, ultra thick] (-0.2, 0.2) -- (0.2, -0.2);
	\node  [draw] at (3,1.5) {C: $c^N$};
	\node  at (5.0,3.7) {D: $c^N$ (?)};
	\node [text width=2.7cm] at (2.5,4.5) {E: unstable (?)};
	\node [red] at (-1.0, -0.0) {B: $N^2$};
	\node [color=black!30!green] at (4.7, -0.4) {A: $N$};
\end{tikzpicture}
\fi
\caption{Disturbance scaling with respect to string length $N$ for different selections of asymmetry. Area A: linear scaling in $N$, Area B: quadratic in $N$, Area C and D: exponential in $N$. {The behaviour in regions marked (?) is conjectured.}}
\label{fig:completePicture} 
\end{figure}
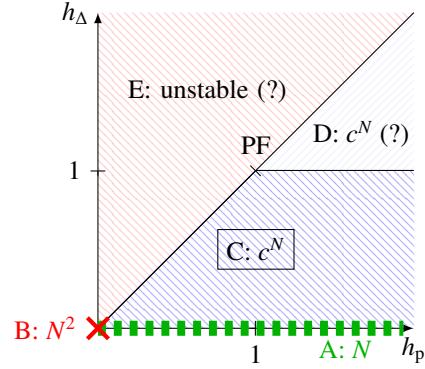

{Figure~\ref{fig:completePicture} summarises how the scaling factor $\xi$ scales with $N$ for different choices of $\hp$ and $\hd$.} The point $\hp=\hd=0$ corresponds to the symmetric, bidirectional case while $\hp=\hd=1$ describes the unidirectional (``predecessor following'' = `PF') case.
The findings can be summarised as follows:
\begin{enumerate}
	\item It is shown that asymmetry in velocity with symmetric position coupling, i.e., case A in Fig.~\ref{fig:completePicture}, achieves linear scaling, while a completely symmetric control scales quadratically, i.e., case B in Fig.~\ref{fig:completePicture}. See Section~\ref{sec:symPosCoup}.\footnote{To the best knowledge of the authors, this is the first paper which analytically proves better scaling when symmetry in position and asymmetry in velocity is used. The papers \cite{Cantos2014a, Herman2016} relied in their proofs only on (reasonable, though) conjectures.}
	\item It is shown that for asymmetric position coupling below a certain bound, the errors scale exponentially, i.e., case C in Fig.~\ref{fig:completePicture}, see Section~\ref{sec:APstab}. We conjecture that it is also true for $\hd\geq1$, i.e., case D in Fig.~\ref{fig:completePicture}.
	\item It is shown that for some cases of stronger asymmetry in the position coupling compared to the velocity coupling, i.e., for a subset of $\hd>\hp$ of case E in Fig.~\ref{fig:completePicture}, the system is unstable for a sufficiently high $N$, see Section~\ref{sec:APunstab}. We also conjecture that for all combinations captured in case E a finite critical string length, which is the maximal stable string length, exists. 
	A discussion of the results is found in Section~\ref{sec:discussion}.
	\item Extensive numerical results are presented to illustrate the technical results, see Section~\ref{sec_example}.
	\item A comprehensive overview of the effect of asymmetry in velocity and position coupling is given. The system description unifies several existing well studied platoon descriptions such as unidirectional strings in \cite{Seiler2004a}, bidirectional symmetric strings as in \cite{Knorn2014,Barooah2005} and bidirectional asymmetric strings as in \cite{Barooah2009a, Cantos2014a}.
	\end{enumerate} 
{\bf Notation}: The $L_2$ vector norm is $|x|_2 = \sqrt{x^\mathrm Tx}$ and the $L_2$ vector function norm $\|x(\cdot)\|_2 = \sqrt{\int_0^\infty |x(t)|_2^2\d t}$. For a scalar function $H(x)$ of a vector $x = \left[x_1, x_2, \dots , x_n \right]^\mathrm T$ its gradient is defined as $\nabla H(x) = \left[ \frac{\partial H(x)}{\partial x_1}, \frac{\partial H(x)}{\partial x_2}, \dots ,  \frac{\partial H(x)}{\partial x_n} \right]^\mathrm T$. The $i$th element of the gradient $\frac{\partial H(x)}{\partial x_i}$ is also denoted $\nabla_{x_i} H(x)$. 
The column vector with $N$ elements is denoted $x(t)= \mathrm{col} (x_1(t),\dots,x_N(t))$. 
{The column vector of ones of length $N$ is denoted by $\underline 1$} and $\vec e_i$ is the $i$th canonical vector of length $N$. Denote the diagonal matrix $A\in\mathbb R^{N\times N}$ with diagonal entries $a_1,\dots a_N$ as $A = \mathrm{diag}(a_1,\dots a_N)$. {The matrix $\langle A \rangle$  is a matrix obtained from $A$ by taking the absolute values of the elements.} $A>0$ and $A\geq0$ denote that $A$ is a positive definite or positive semi-definite matrix, respectively. $\sigma_i(A)$ is the $i$th smallest singular value of $A$ and $\lambda_i(A)$ is the $i$th smallest eigenvalue. $\smin(A)$, $\smax(A)$ ($\lmin(A), \lmax(A)$) are the minimal and maximal singular values (eigenvalues) of $A$, respectively.

\section{Mathematical preliminaries}\label{sec_prel}

Consider a string of $N$ vehicles with dynamics \eqref{eq_xddot} and \eqref{eq_Fi}. It is assumed that the control parameters and vehicle masses are constant and bounded from below and above:
\begin{ass}\label{ass_r}
	{There exist constants $\underline r>0$, $\overline r<\infty, \,\, \underline m > 0, \overline m < \infty, \:\: \underline a > 0, \overline a < \infty$ such that $\underline r\leq r_i \leq \overline r$, $\underline m \leq m_i \leq \overline m$ and $\underline a \leq a_i \leq \overline a$ for all $i\leq N$ and for all $N$. }
	\label{ass:constants}
	\end{ass}

{By collecting the positions in $x(t) = \mathrm{col}(x_1,\dots,x_N)$, and introducing $\Delta(t) = \mathrm{col}(\Delta_1,\dots,\Delta_N)$, $\dreference = \mathrm{col}(\dreference_{0, 1}, \ldots, \dreference_{0, N})$ the local position errors can be represented by
\begin{equation}
	\Delta(t) = -\BBT(x(t)-\underline 1 x_0(t)+\dreference), 
	\end{equation}}
where $\BB$ describes the coupling between the vehicles
\begingroup
\begin{equation}
	\BB = \begin{bmatrix}1 & -1  & 0 &  \cdots & 0\\
	0 & \ddots &\ddots & \ddots &0 \\
	\vdots &  & \ddots & 1 &-1 \\
	0 & \cdots & 0 & 0 &1\end{bmatrix}.
	\label{eq:bbDelta}
	\end{equation}
\endgroup
Let $R=\mathrm{diag}(r_1,\dots,r_N)>0$ be the matrix of damping coefficients and $M = \mathrm{diag}(m_1,\dots,m_N)>0$ be the inertia matrix. Further, let the velocity asymmetry matrix be given as
%
	$\BBpt = \hp\Babs.$ 
 Then the vector of forces due to relative velocities is described by
\begin{equation}
	F^\mathrm r = -(\BB+\BBpt) R\BBT(M^{-1}p-\underline 1v_0), \label{eq:dampingForces}
	\end{equation}
where $p = \mathrm{col}(p_1,\dots,p_N)\in \mathbb{R}^N$ is the momentum vector. 
%
%
The vector of forces due to position errors can be written as 
\begin{equation}
		F^\mathrm s = (\BB + \BBdt) A\Delta,
	\end{equation}
where $A=\mathrm{diag}(a_1,\dots a_N)>0$ and $\BBdt=\hd \langle \BB \rangle$. Hence, the vector of control forces $F = \mathrm{col}(F_1,\dots,F_N)\in \mathbb{R}^N$ is $F=F^\mathrm r+F^\mathrm s$.

{We will analyse the following asymmetry combinations:
\begin{itemize}
	\item $\hd = 0, \hp=0$, which refers to symmetric position and symmetric velocity coupling, abbreviated as \emph{SPSV}.
	\item $\hd = 0, \hp>0$, which refers to symmetric position and asymmetric velocity coupling, abbreviated as \emph{SPAV}. 
	\item $\hd > 0, \hp\geq0$, which refers to asymmetric position and asymmetric velocity coupling, abbreviated as \emph{APAV}.
\end{itemize}}


\section{Symmetric position coupling}\label{sec:symPosCoup}

{In this section we investigate the cases with symmetric spring forces, \ie, $\hd = 0$ (and thus $\BBdt=0$), such that the platoon can be written in the port-Hamiltonian form}
\begin{equation}\label{eq:clphSym}
	\begin{bmatrix} \dot{p}(t) \\ \dot{\Delta}(t) \end{bmatrix}
	= \begin{bmatrix} -(\BB+\BBpt) R \BBT & \BB \\ 
		-\BBT & 0\end{bmatrix} \nabla H(p(t),\Delta(t)) + \begin{pmatrix} d(t) \\ 0 \end{pmatrix},
	\end{equation}
with the Hamiltonian function
\begin{equation}\label{eq_H}
	H(p,\Delta)
	= \frac{1}{2}(p(t)-M\underline 1v_0)^\mathrm TM^{-1} (p(t)-M\underline 1v_0) +\frac{1}{2} \Delta^\mathrm T(t)A\Delta(t)
	\end{equation}
and the equilibrium
\begin{equation}\label{eq:equilibrium}
	\Delta_i = 0 \text{ and } v_i = v_0 \text{ for all } i\leq N.
	\end{equation}
The Hamiltonian $H(p,\Delta)$ captures the kinetic ``energy'' stored in the relative velocity to the leader and potential ``energy'' stored in the position errors. {Hence, it captures both the spacing and velocity errors of all vehicles. In the following we provide an upper bound on the Hamiltonian, leading to a bound on the maximal error. The following assumption is necessary to guarantee, that \eqref{eq_H} is a suitable Hamiltonian function for SPAV systems.}
\begin{ass}\label{ass}
	$r_i \geq r_{i+1}$ for all $i<N$.
\end{ass}

{SPSV and SPAV systems have similar properties:}

\begin{thm}[SPSV, SPAV]{
	Consider the system \eqref{eq:clphSym} under Assumption \ref{ass:constants} with Hamiltonian \eqref{eq_H}. 
	If $\hp > 0$ (SPAV),  the following holds only under Assumption \ref{ass}; if $\hp=0$ (SPSV), it holds unconditionally:	%
	\renewcommand{\theenumi}{\roman{enumi}}
	\begin{enumerate}
	  \item for $d(t)\equiv 0$ the equilibrium \eqref{eq:equilibrium} of the autonomous system is asymptotically stable, 
	  \item the system is passive with input vector $d(t)$, output vector $\nabla_p H$ and storage function \eqref{eq_H},
	  \item the following bound on the Hamiltonian at time $t$ holds
	  \begin{equation}
	  	H(p(t), \Delta(t)) \leq H(p(0), \Delta(0)) + \frac{\|d(\cdot)\|_2^2}{2 \sigma_{\min}\left((\BB+\hp \langle \BB\rangle ) R \BBT\right)}. \label{eq:boundScSPSV}
	  \end{equation}
	\end{enumerate}}
	
	\label{thm:SPXV}
\end{thm}

\begin{pf}
{The proof for the case $\hp=0$ is presented in \cite{Knorn2016}. Thus, only the proof for $\hp>0$ is presented here.}
	
	(i) Take $H$ in \eqref{eq_H} as a Lyapunov function and set $d(t)=0$. The time derivative of $H$ is
	\begin{equation}
		\dot H = \nabla^\T H  \begin{bmatrix} -\left(\BB+\BBpt\right) R \BBT & \BB \\ 
		-\BBT & 0\end{bmatrix} \nabla H.
		\end{equation}
	This yields $\dot H = -\nabla_p^\T H \left(\BB+\BBpt\right) R \BBT\nabla_p H$. In order to show that $-\nabla_p^\T H \left(\BB+\BBpt\right) R \BBT \nabla_p H<0$, let {$\vt_i$} be the $i$th element of $\nabla_p H$. Then, since $\left(\BB+\BBpt\right) R \BBT$ has the tridiagonal structure (it is a graph Laplacian of the path graph)
\begingroup
	\begin{equation}
		\begin{bmatrix}
			(1+\!\hp)r_1\!+\!(1\!-\!\hp)r_2 & -(1\!-\!\hp)r_2 &  & 0 \\
			-(1+\hp)r_2 & \ddots & \ddots &   \\
			  & \ddots & \ddots &  -\!(1\!-\!\hp)r_N\\
			0 &  & \!-(1\!+\!\hp)r_N & (1\!+\!\hp)r_N
			\end{bmatrix},
			\label{eq:laplAsymVel}
	\end{equation}
\endgroup
	it follows that
	\begin{align}
		\nonumber &-\nabla_p^\T H \big(\BB + \BBpt\big)R\BBT \nabla_p H\\
		\nonumber =& -\left(r_1(1+\hp)+r_2(1-\hp) \right) \vt_1^2 + r_2(1-\hp)\vt_1 \vt_2 \nonumber \\
		&+ r_2(1+\hp) \vt_2 \vt_1- \left(r_2(1+\hp) + r_3(1-\hp)\right) \vt_2^2 + \ldots \nonumber\\
		 \nonumber&- r_N(1+\hp)\vt_N^2 \\
		 \nonumber \leq& -r_1 \vt_1^2- r_2(\vt_1-\vt_2)^2 - r_3(\vt_2-\vt_3)^2 - \dots \nonumber \\
		 \nonumber& - r_{N-1}(\vt_{N-1}-\vt_{N})^2-r_N \hp \vt_N^2
		 < 0,
		\end{align}
	where in the first inequality we used Assumption~\ref{ass}. Hence, the system is Lyapunov stable. {Asymptotic stability follows using the invariance principle, see \cite{Khalil}. The set when $\dot H = 0$ is $\nabla_\Delta H = \BB A \Delta$. Since $A$ and $\BB$ are nonsingular it follows that the only positively invariant set is $\Delta = 0$}.
	
	(ii) Considering $d(t)$, the derivative of the Lyapunov function \eqref{eq_H} is $\dot H = -\nabla_p^\T H \left(\BB + \BBpt\right) R \BBT \nabla_p H + \nabla_p^\T H d(t)$. Taking $y=\nabla_p H$ as an output yields
	\begin{equation}
		\dot H \leq -\smin\left(\left(\BB + \BBpt\right)R \BBT\right)|y|_2^2 + y^\T d.  \label{eq:dotHsigma}
		\end{equation}
	The increase in the energy of the system $H$ is less than the ``power'' $y^\T d$ applied to the system. The system is passive.
	
	(iii) Extending \eqref{eq:dotHsigma} by completing the squares leads to
	\begin{align}
		\nonumber \dot H \leq& -\frac{\smin\left(\left(\BB + \BBpt\right)R\BBT\right)}{2}|y|_2^2 + \frac{|d(t)|_2^2}{2 \smin \left(\left(\BB + \BBpt\right)R\BBT\right)}
		\\ \nonumber & -\frac{\smin\left(\left(\BB + \BBpt\right)R\BBT\right)}{2} \left| y-\frac{d(t)}{\smin \left(\left(\BB + \BBpt\right)R\BBT\right)}\right|_2^2 \\
		 \leq  & \frac{|d(t)|_2^2}{2 \smin \left(\left(\BB + \BBpt\right)R\BBT\right)}. 
		\label{eq_bound_D}
		\end{align}
	Integrating with respect to $t$ yields \eqref{eq:boundScSPSV}. $\hfill\strut$\qed
\end{pf}

\subsection{Scaling of singular values}\label{sec:singvalue}
{
Suppose that the norm of the disturbance signal is fixed for any $N$, \ie,  $\|d(\cdot)\|_2=\mathrm{const.}$ Then in Theorem \ref{thm:SPXV}, the effect of the disturbance depends on the minimal singular value of the damping matrix $(\BB+\hp\langle \BB\rangle) R \BBT$. The smaller $\smin\left((\BB+\hp \Babs)R\BBT\right)$ is, the larger will be the effect of the disturbance $d(t)$ on the total energy of the system, and therefore, on the deviations from the equilibrium. The worst-case convergence rate \eqref{eq:dotHsigma} also depends on this singular value. } 

Next, we will investigate how the smallest singular value scales with $N$. First, we consider the lower bound on the smallest singular value. For the proof see Appendix~\ref{app:lemLowerBound}.
\begin{lem}
	{Let $\gamma=\min\{1, 1/\hp\}$. Then,
	\begin{equation}
		\smin \left((\BB+\hp\Babs) R \BBT \right) \geq \underline r \sqrt{\frac{1}{16}\frac{1}{N^4} + h_p^2 \gamma \frac{1}{N^2}}.
		\label{eq:scalingLowerBound}
	\end{equation}}
	\label{lem:scalingLowerBound}
\end{lem}
The final scaling result, proven in Appendix~\ref{app:lemScalingSigma}, is as follows.
%
\begin{lem}
	{With some positive constants $c_1, c_2, c_3, c_4$ and for $N$ sufficiently large,
	\begin{align}		
		\frac{c_1 \, \underline r}{N} & \leq \smin \left(\left(\BB+\hp \Babs\right) R \BBT\right) \leq \frac{c_2 \, \overline r}{N}	  
			 \quad \text{for } \hp > 0, \label{eq:scalingAsym}
			 \\
		\frac{c_3 \,\underline r}{N^2} & \leq \smin \left(\left(\BB+\hp \Babs\right) R \BBT\right) \leq \frac{c_4 \,\overline r}{N^2}
			\quad \text{for } \hp = 0. \label{eq:scalingSym}
	\end{align}}
	\label{lem:scalingSigma} 
	\end{lem}

{From Lemma~\ref{lem:scalingSigma} note that, the upper and lower bounds on $\smin$ are of the same order and approach zero as the number of vehicles grows} both for SPAV and SPSV. The rate of approach is quadratic for SPSV, while it slower -- linear -- for SPAV. This means that when asymmetric coupling in velocity is used, the effect of the disturbance is qualitatively smaller than in SPSV systems. 

\begin{rem}{Consider that $\hp$ is very small such that $\gamma=1$. Then, for sufficiently small $N$, $\frac{1/16}{N^4} \gg \hp^2 \frac{1}{N^2}$ such that the singular value scales quadratically. However, if
\begin{equation}
	N \geq 1 / (4\hp).
	 \label{eq:minNLin} 
	\end{equation}
the second term in the square root in (\ref{eq:scalingLowerBound}) becomes greater than the first term and the scaling improves to linear.}
\end{rem}

\begin{figure}
\centering 
	\includegraphics[width=0.40\textwidth]{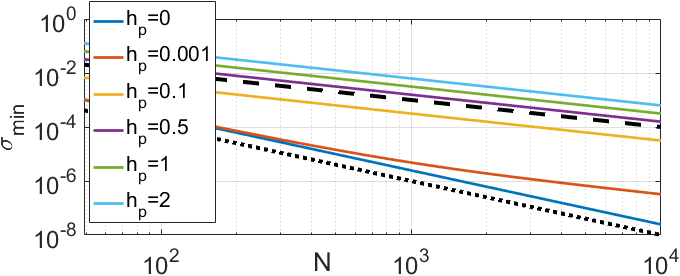}
	\caption{Scaling of $\sigma_{\min}\left(\left(\BB+\hp \Babs\right)R\BBT\right)$ for $R=I$ as a function of $N$ with $\hp=0$ (SPSV) and various $\hp>0$ (SPAV). The dotted line is $1/N^2$ and the dashed line is $1/N$.}
	\label{fig:scalingSigma}
\end{figure}

These results are also verified numerically. The scaling of the smallest singular value of the velocity coupling matrix $\left(\BB+\hp \Babs\right)R \BBT$ is shown in Fig. \ref{fig:scalingSigma}. It is clear that asymmetric coupling achieves scaling with rate $1/N$, while symmetric coupling approaches zero faster, \ie, as $1/N^2$. Note that the larger the asymmetry (greater $\hp$), the larger also the smallest singular value.\footnote{Time-domain plots and scaling of other important quantities are illustrated in the Sec. \ref{sec_example}. It confirms that in any of the quantities SPAV achieves better results.} {Also note that for very small asymmetry $\hp=0.001$, the scaling was quadratic for low $N$, while it improved to linear for $N>250$ as expected, \eqref{eq:minNLin}.}

\begin{rem}{Lemma \ref{lem:scalingLowerBound} together with Fig. \ref{fig:scalingSigma} suggest that $\smin \big((\BB+\BBpt)R\BBT\big)$ increases with $\hp$, leading to smaller deviation bounds and faster convergence rates. Although this is mathematically true, a practical implementation with $\hp > 1$ might be fragile as the coupling of the vehicle $i$ with vehicle $i+1$ gets a positive sign, causing a positive feedback. While stability of the formation is guaranteed by the strong coupling with vehicle $i-1$, stability might be lost in case of erroneous relative velocity measurements. In addition, $\hp\gg1$ also leads to high gains and potentially to actuator saturation. Hence, setting $\hp \leq 1$ is a better solution.}
\end{rem}
 

\section{Asymmetric position coupling}\label{sec:AP}

{In this section we allow that also the coupling in position is asymmetric, hence $\hd \geq 0$. The discussion here is divided in two parts based on the ratio of $\hp$ and $\hd$.}

\subsection{Asymmetry in position less than in velocity, $\hd\leq\hp$}\label{sec:APstab}

In case $\hd < 1$, the overall platoon can be modelled in the port-Hamiltonian form
\begin{equation}
		\begin{bmatrix} \dot{p}(t) \\ \dot{\Delta}(t) \end{bmatrix}
		\!=\! \begin{bmatrix} - \left(\BB\!+\!\BBpt\right) R \BBT E^{-1} & \frac{1}{1+\hd}(\BB\!+\!\BBdt)E^{-1} \\  -\BBT E^{-1} & 0 \end{bmatrix}	
			 \nabla H_\Delta
		 	+ \begin{bmatrix}  d(t) \\ 0\end{bmatrix}. 
		\label{eq:portHamAsymPos}
		\end{equation}
where
\begin{equation}
\begin{aligned}\label{eq_Hdelta}
	H_\Delta(p,\Delta)
	=& \frac{1}{2}(p(t)-M\underline 1v_0)^\T E M^{-1} (p(t)-M\underline 1v_0)  \\ 
	&+\frac{1}{2} \Delta^\T (t) (1+\hd)EA\Delta(t).
	\end{aligned}
\end{equation}
with $E = \mathrm{diag}\left(1,\frac{1-\hd}{1+\hd},\left(\frac{1-\hd}{1+\hd}\right)^2,\ldots, \left(\frac{1-\hd}{1+\hd}\right)^{N-1}\right)$ being the scaling matrix.
%
%
	Then, in {Appendix \ref{app:APAV}} we prove the following properties.

\begin{thm}[APAV]
	Consider system \eqref{eq:portHamAsymPos} under Assumptions~\ref{ass:constants} and \ref{ass} with Hamiltonian \eqref{eq_Hdelta}. Then,
	\begin{itemize}
	  \item[(i)] the equilibrium \eqref{eq:equilibrium} of the autonomous system is asymptotically stable for $\hp
	\geq \hd$ and $\hd < 1$, and
	\item[(ii)] the effect of disturbances is bounded as
		\begin{equation}
			H_\Delta(t) \leq H_{\Delta}(0) +\frac{\|d (\cdot)\|_2^2}{2 \sigma_{\min}\left(E\left(\BB+\BBpt\right) R \BBT\right)}, \label{eq:distBoundAPAV} 
			\end{equation} 
		 and $\sigma_{\min}\left(E \left(\BB+\BBpt\right) R \BBT \right)$ for $N$ sufficiently large approaches zero with rate $1/c^N$, with $c>1$.
		\end{itemize}
	\label{thm:expScaling}	
	\end{thm}

Theorem~\ref{thm:expScaling} means that the upper bound on the effect of the disturbance scales exponentially in the number of vehicles, which is qualitatively much worse than the scaling for symmetry in position coupling. So breaking up the symmetry in position significantly deteriorates the performance.

\begin{rem}
 	{Theorem~\ref{thm:expScaling} above yields an upper bound. Although this does not mean that the system does not scale better, the results in the literature confirm that at least for a particular cases where $\hp=\hd$ the $\hinfnorm$ norm of any transfer function in the formation scales exponentially with the graph distance \cite{Herman2014b}.  Similar results appeared in \cite{Seiler2004a, Tangerman2012, Herman2013b}.} 
\end{rem} 

\begin{rem}{For the transition between APAV and SPAV (or SPSV) consider $\hd$ to be very small and the number of vehicles sufficiently low. Then the scaling matrix $E \approx I$, hence also its smallest singular value is very close to one. Then $H_\Delta \approx H$, $\frac{\|d(\cdot)\|_2}{2 \smin\left(E (\BB+\BBpt)R\BBT\right)} \approx \frac{\|d(\cdot)\|_2}{2 \smin\left((\BB+\BBpt)R\BBT\right)}$ and \eqref{eq:distBoundAPAV} becomes \eqref{eq:boundScSPSV}. Thus, the scaling is similar to the scaling of SPAV or SPSV, depending on the parameter $\hp$. But for $N$ large enough, the scaling becomes exponential.}
\end{rem}

\subsection{Asymmetry in position greater than in velocity, $\hd>\hp$}\label{sec:APunstab}

It will be shown that even short strings of length $N=2$ become unstable for particular combinations of $\hp$ and $\hd$.
For simplicity, set $R=A=M=I$ and $\dreference=0$. Then, the dynamics of vehicle $i$ for $i<N$ are given by
\begin{equation}
\begin{aligned}
	\ddot x_i =& (1+\hp)(v_{i-1}-v_i) + (1-\hp)(v_{i+1}-v_i)\\
		& +(1+\hd)(x_{i-1}-x_i) + (1-\hd)(x_{i+1}-x_i),
	\end{aligned}
\end{equation}
whereas the last vehicle is described by
%
	$\ddot x_N = (1+\hp)(v_{N-1}-v_N) +(1+\hd)(x_{N-1}-x_N).$
	%
Applying the Laplace transform to both equations above and denoting $X_i(s)=\mathcal L\{x_i(t)\}$, or just $X_i$, leads to the following transfer function relations
\begin{align}
	 X_i\! =&\! \underbrace{\frac{(1\!+\!\hp)s\!+\!(1\!+\!\hd)}{s^2+2s+2}}_{:=G^-}X_{i-1}
		\label{eq:Xis} + \underbrace{\frac{(1\!-\!\hp)s\!+\!(1\!-\!\hd)}{s^2+2s+2}}_{:=G^+}X_{i+1}\\
	\label{eq:XNs} X_N =& \underbrace{\frac{(1+\hp)s+(1+\hd)}{s^2+(1+\hp)s+1+\hd}}_{:=G_N}X_{N-1}
	\end{align}
	\begin{figure}
\centering
\iftikz
\begin{tikzpicture}[auto, >=latex]
	\node[placeholder] (inp) {$$}; 
	\node[block]  (t1) [right=of inp, right=2.5em] {$G_1$};
	\draw [->] (inp) -- node[midway] {$X_0$} (t1);
	\node  (dots1)	[right=of t1, right=2.5em] {$\ldots$};
	\draw [->] (t1) -- node [midway]{$X_1$} (dots1);
	\node[block]  (t2) [right=of dots1, right=2.5em] 	{$G_{N-1}$};
	\draw [->] (dots1) -- node [midway] {$X_{N-2}$} (t2);
	\node[block]  (z1) [right=of t2,	right=2.5em]	{$G_N$}; 
	\draw [->] (t2) -- node [midway]  {$X_{N-1}$} (z1);
	\node[placeholder] (out) [right=of z1, right=2.5em] {$$};
	\draw [->] (z1) -- node[midway] {$X_N$} (out) ;
\end{tikzpicture}
\fi
\caption{String model using the transfer function description $X_i(s) = G_i(s)X_{i-1}(s)$ with \eqref{eq:Xis}-\eqref{eq:XNs}.}
\label{fig:stringtransferfunctions}
\end{figure}
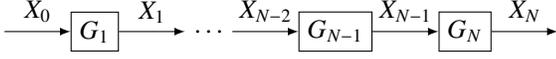
Then, by writing $X_{i}(s)=G_i(s)X_{i-1}(s)$, the transfer functions $G_i$ for $i<N$ can be derived recursively using the relation $G_i = (1-G^+G_{i+1})^{-1}G^-$ such that the dynamics of the entire string can be described as illustrated in Fig.~\ref{fig:stringtransferfunctions}.
Then, it can be shown that for specific choices of $\hp$ and $\hd$, strings of length $N=2$ are unstable\footnote{Note that $G_i = (1-G^+G_{i+1})^{-1}G^-$ is equivalent to a positive feedback loop of BIBO stable subsystems, which is known to potentially lead to closed-loop unstable systems.}:

\begin{lem}\label{lem:instability}
	Strings of $N\geq2$ vehicles are unstable if $\hd\geq \frac{2\hp^4 + 12\hp^3 + 25\hp^2 + 30\hp + 11}{3\hp^2 + 6\hp + 7}$ or $\hd\geq \frac{\hp^3+5\hp^2 +8\hp+10}{\hp-1}$ for $\hp>1$.
	\end{lem} 
 
\begin{pf}
	The results in \eqref{eq:Xis} and \eqref{eq:XNs} and tedious calculations reveal that the denominator of $G_{N-1}$ is given by $\mathrm{den}_{N-1} = s^4+(3+\hp)s^3 +(3+\hd+(1+\hp)^2)s^2 +2(1+\hp)(1+\hd)s+(1+\hd)^2$. Using the Hurwitz stability criterion, it is evident that all roots of $\mathrm{den}_{N-1}$ have negative real parts if and only if $\hd>-1$, $\hp>-3$, $\hd\leq \frac{2\hp^4 + 12\hp^3 + 25\hp^2 + 30\hp + 11}{3\hp^2 + 6\hp + 7}$ and $\hd\leq \frac{\hp^3+5\hp^2 +8\hp+10}{\hp-1}$ in case $\hp>1$. Hence, if one or more of those bounds are violated, then the second last transfer function in the string is unstable, leading to an overall unstable string. Since only nonnegative $\hp$ and $\hd$ are considered here, the bounds yield the result. $\hfill\strut$\qed
	\end{pf}

\begin{figure} 
\centering
	\includegraphics[width=0.25\textwidth]{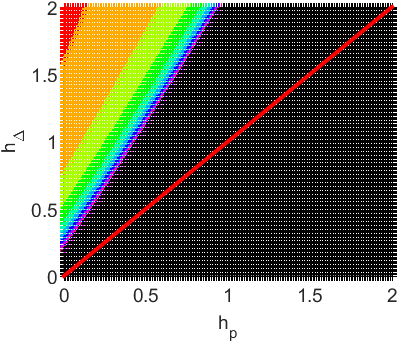}
	\caption{Maximal stable string length $\bar N_\mathrm{stab}$ as a function of $\hp$ and $\hd$: $\bar N_\mathrm{stab} = 1$ in red, $\bar N_\mathrm{stab} = 2$ in orange, \dots, $\bar N_\mathrm{stab} =10$ in purple, $\bar N_\mathrm{stab} > 10$ in black}
	\label{fig:stabborder}
\end{figure}

	Similar bounds on $\hd$ can also be found for the stability of the third last vehicle in the string using the same method as in the proof of Lemma~\ref{lem:instability}. Fig.~\ref{fig:stabborder} illustrates the maximal, stable string length $\bar N_\mathrm{stab}$ for combinations of $\hp$ and $\hd$. It can be seen that longer strings remain stable for smaller ratios $\hd/\hp$. Also, from Theorem \ref{thm:expScaling} it follows that strings of arbitrary lengths are stable if $\hd\leq\hp$ and $\hd < 1$ (where the border of this region is marked with a red, solid line). Based on those observations, we formulate the following conjecture.

\begin{con}
	For every $\hd>\hp$ there exists a maximal string length $\bar N_\mathrm{stab}$ for which the system is stable and all strings of length $N>\bar N_\mathrm{stab}$ are unstable.
	\label{con:instab}
	\end{con}





%


\section{Examples and simulations}\label{sec_example} 

First, illustrative simulations confirming the quadratic and linear growth are discussed. The simulation setup is the following. 
The input signal is $d_i(t)={p_i}(t)/\left({\|p\|_2 \sqrt{\Tf}}\right)$ for $t<\Tf$ and $d_i(t)=0$ for $t>\Tf$. That is, the disturbance vector is parallel to the momentum vector, which by \eqref{eq:dotHsigma} achieves the fastest growth of the energy. The time $\Tf$ is some given duration of the signal. The $L_2$ norm $\|d(\cdot)\|_2$ is then $1$ for all $N$. All the simulations were started with zero initial conditions, hence $H(0)=0$ in \eqref{eq_H}. As follows from our theorems, we are interested in the maximal value of the Hamiltonian functions $H_{\max}=\max_{t} H(p(t), \Delta(t))$.  

\begin{figure} 
\centering
	\subfloat[SPSV]{\includegraphics[width=0.24\textwidth]{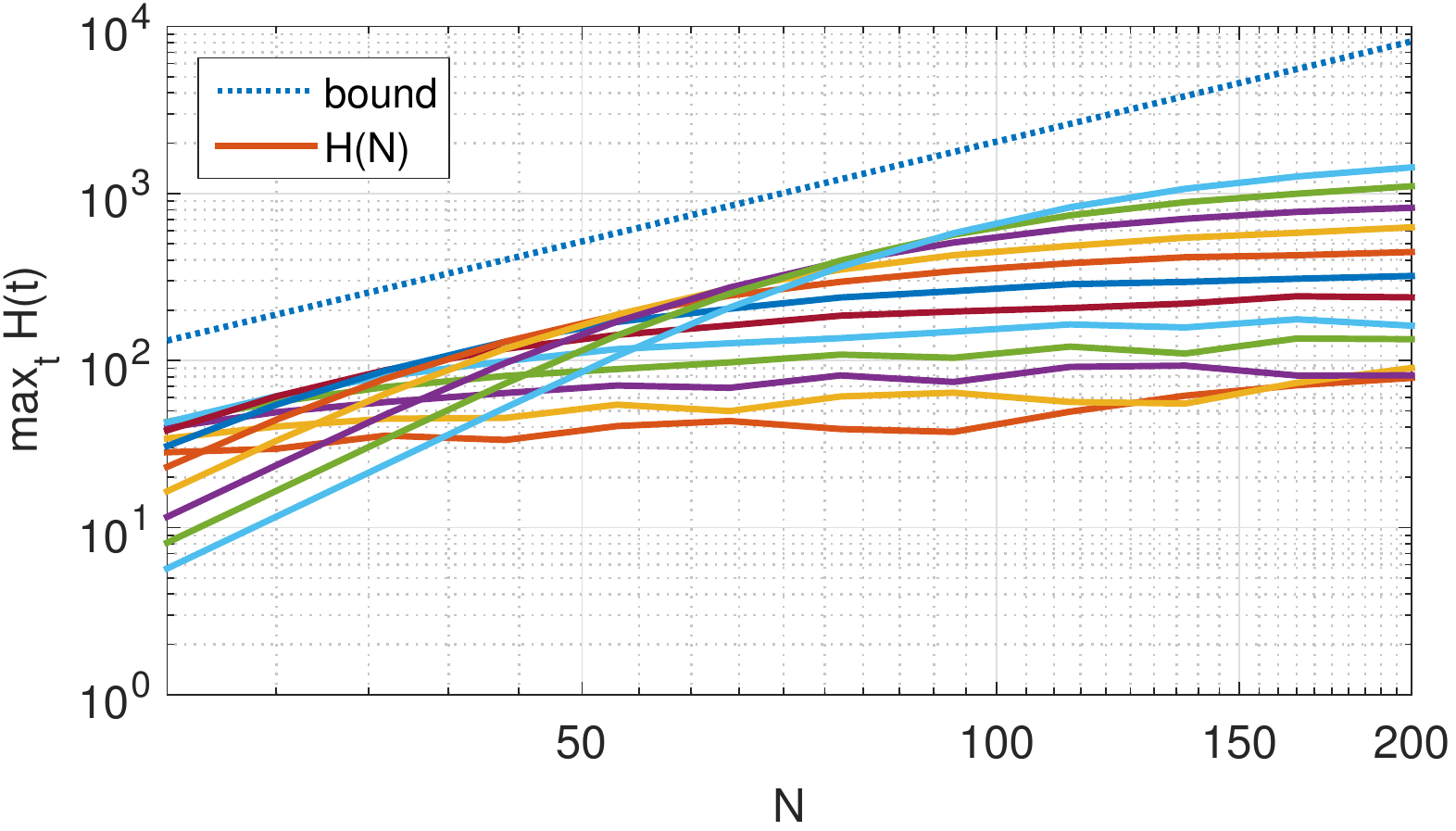}  \label{fig:maxEnGrowSPSV}}
	\subfloat[SPAV]{\includegraphics[width=0.24\textwidth]{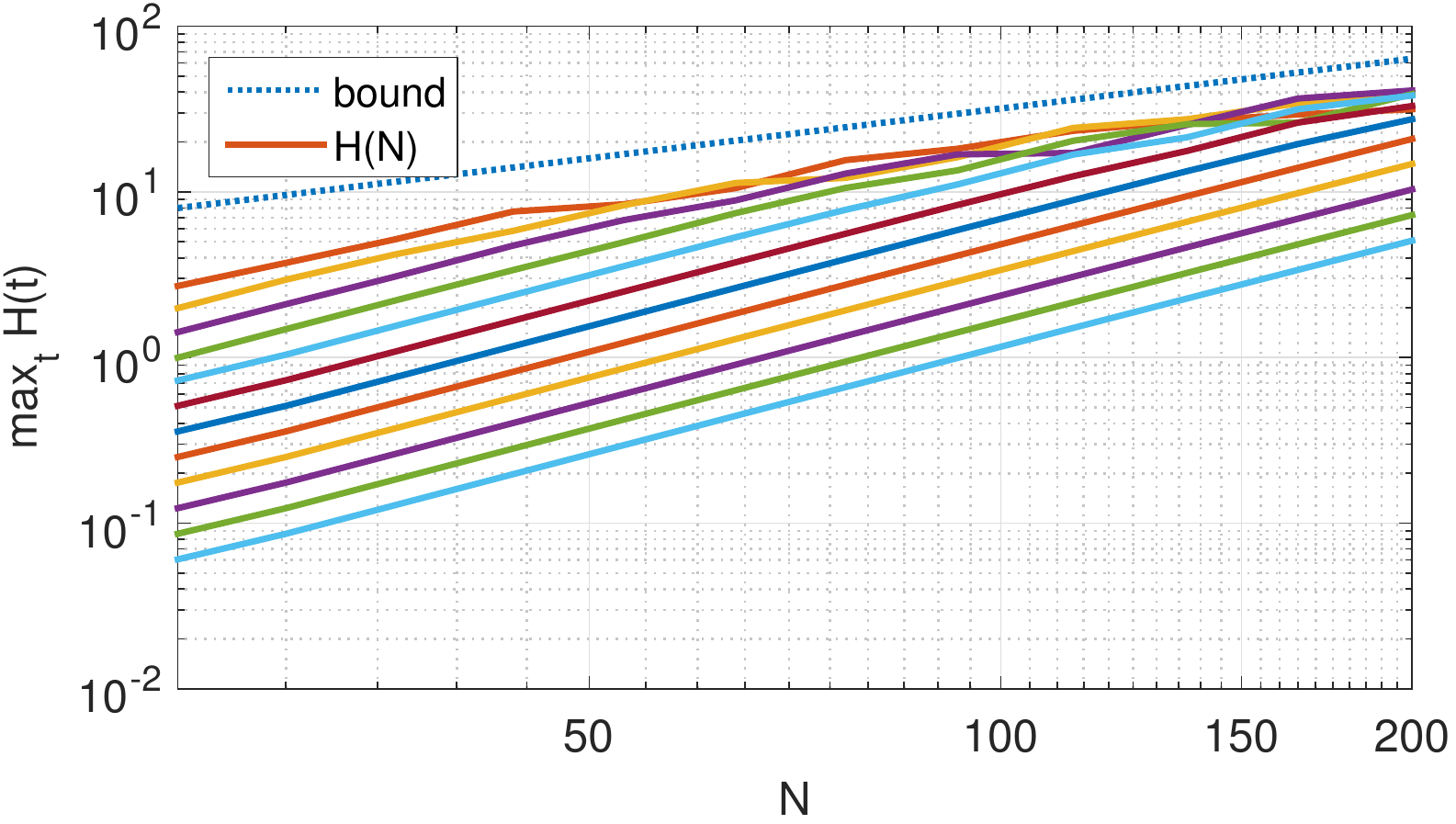} \label{fig:maxEnGrowSPAV}}
	\caption{Plot of maximal value of $H(t)$ for SPSV (a) and SPAV (b) in logarithmic coordinates when the input is applied for different times ($\Tf \in [200, 10000]$) and $N=25, \ldots, 200$. Dashed line is the bound \eqref{eq:boundScSPSV}.}
	\label{fig:maxEnGrow}  
\end{figure}

First consider SPSV. Different durations of the input signal in the range $\Tf \in [200, 10000]$ were used. The plot of $H_{\max}$ is shown in Fig. \ref{fig:maxEnGrow}a for SPSV. The solid lines correspond to different $\Tf$. Note that, the longer the duration $\Tf$ is chosen, the lower the value for low $N$ and the higher the maximal value of $H$ become. On the curve for each $\Tf$, consider the point which is the closest to the bound. From the plot in Fig. \ref{fig:maxEnGrowSPSV} it is apparent that this point scales quadratically with $N$. It means that the quadratic growth was achieved. The case with SPAV behaves similarly. The growth of the point, where each curve gets closest to the bound, is linear. This is illustrated in Fig. \ref{fig:maxEnGrow}b. Thus, the results of Theorem \ref{thm:scalingAll} were verified. Also note that, for a given $N$, $\max_t H(t)$ is much smaller for case SPAV, than for SPSV. Although the bounds are conservative, they capture the scaling qualitatively.

\subsection{Characteristics of the transient} 
\begin{figure}
\centering
		\subfloat[Momentum SPSV, SPAV]{\label{fig:timePlotVel}\includegraphics[width=0.22\textwidth]{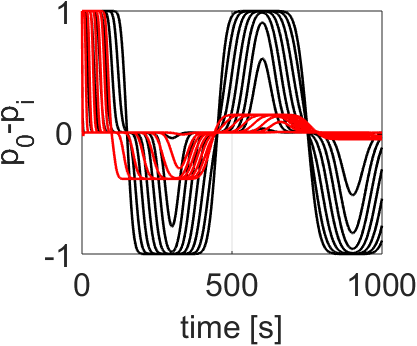}  }
		\subfloat[Pos. error SPSV, SPAV]{\label{fig:timePlotDist}\includegraphics[width=0.22\textwidth]{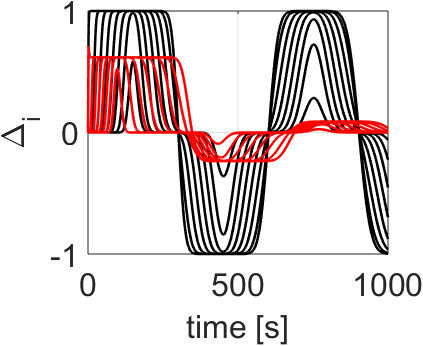} }\\
		\subfloat[Momentum APAV]{\label{fig:timePlotVelAPAV}\includegraphics[width=0.22\textwidth]{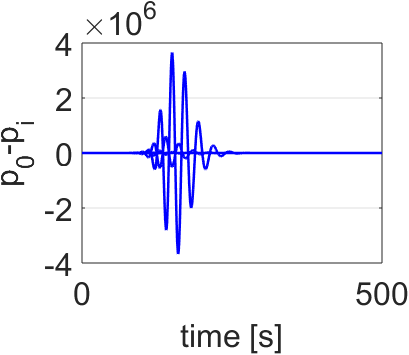}  }
		\subfloat[Pos. error APAV]{\label{fig:timePlotDistAPAV}\includegraphics[width=0.22\textwidth]{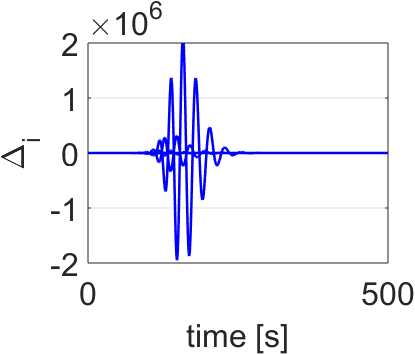} }
	\caption{Response of the platoon with $N=150$ to the leader's step change in velocity. Black is SPSV, red is SPAV, blue is APAV.}
	\label{fig:leaderVelResponse}
\end{figure}
As the test conditions, assume that all the initial states are zero. At time zero, the leader starts to move with unit velocity, so $x_0 = t$. This corresponds to an acceleration manoeuvre of the platoon. Further, it is assumed $r_i=1, a_i=1, m_i=1, \delta_{\text{ref}\,i-1,i}=0 \, \forall i$, in SPAV $\hp=0.5\,\forall i$ and in APAV $\hp=0.5, \hd = 0.2, \, \forall i$.

When one looks at the time-domain plots in Fig. \ref{fig:leaderVelResponse}, it is apparent that SPAV has shorter convergence time and lower overshoots than SPSV. When asymmetry in position is introduced (APAV), extremely high peaks occur. Despite the fact that the leader moves with unit velocity, during the transient the states of some vehicles reached up to the order of $10^6$. 

\begin{figure}
	\centering
	\subfloat[Max. overshoot]{\label{fig:maxOvershoot}\includegraphics[width=0.22\textwidth]{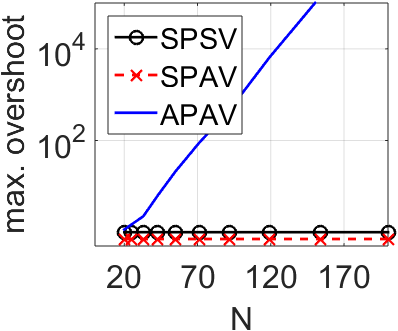}}
	\subfloat[Max. cont. eff.]{\label{fig:controlEffortNrange}\includegraphics[width=0.22\textwidth]{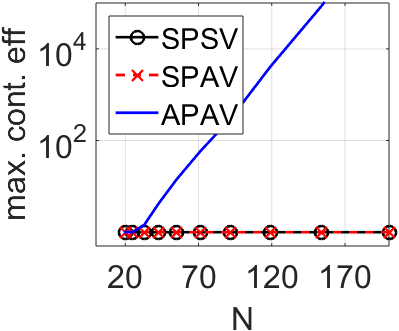}}\\
	\subfloat[Conv. time]{\label{fig:convTime}\includegraphics[width=0.22\textwidth]{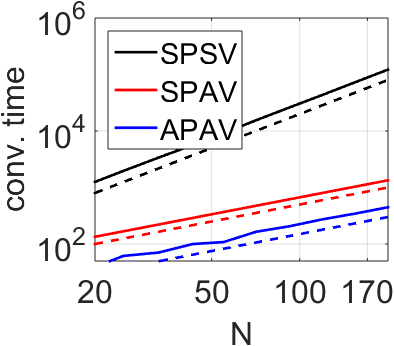}}
	\subfloat[Total error]{\label{fig:totalError}\includegraphics[width=0.22\textwidth]{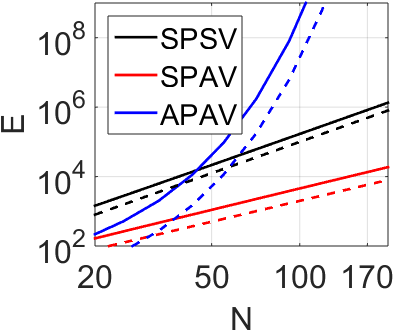}}
	\caption{Scaling of several quantities of interest as a response to the unit step in leader's velocity.  Note that a) and b) are in semilogarithmic coordinates, c) and d) in logarithmic coordinates. In c) the dashed lines are $2N^2$ (black), $5N$ (red) and $1.5N$ (blue). In d) the dashed lines are $0.1N^3$ (black), $0.2N^2$ (red) and $e^{0.17N}$ (blue). }
	\label{fig:scalingOtherQuantities}
\end{figure}

Scaling of other quantities, such as maximal overshoot, maximal control effort, convergence time and total error, is shown in Fig \ref{fig:scalingOtherQuantities}, from which the following can be observed:
\begin{itemize}
  \item \emph{Maximal overshoot} $\max_{i, t} \Delta_i(N,t)$ (Fig. \ref{fig:maxOvershoot}): Both SPSV and SPAV are bounded for all $N$, while SPAV achieves a lower bound. APAV scales exponentially.
  \item \emph{Maximal control effort} $\max_{i, t} F_i(t)$ (Fig. \ref{fig:controlEffortNrange}): Both SPAV and SPSV have the same value equal to one, while APAV scales exponentially. The control effort of SPSV and SPAV does not grow with $N$.
  \item \emph{Convergence time} (Fig. \ref{fig:convTime}): It is apparent that SPAV and APAV scale linearly, while SPSV scales quadratically with $N$. Thus, linear scaling of SPAV and quadratic scaling of SPSV appears also in the convergence time.
  \item \emph{Total error} $E = \sum_{i=1}^{N} \int_0^\infty \Delta_i^2 + (v_i-v_0)^2 dt.$ (Fig. \ref{fig:totalError}): Apparently, SPAV achieves the best scaling, that is, quadratic, SPSV scales cubically and APAV scales again exponentially with $N$. 
\end{itemize}  
 
We conclude that SPAV performs the best in all cases. The only exception is the convergence time. It is true that APAV achieved about 4 times faster transient, but at the price of exponential scaling of any other quantity. We can see that SPAV has similar convergence time as APAV while keeping bounded control effort as SPSV.

\section{Summary and discussion}
\label{sec:discussion}
The results are summarised in the following theorem. 
\begin{thm}
	The qualitative effect of the disturbance on the energy in the system scales with the number of vehicles $N$ as
	\begin{itemize}
		\item (SPSV): $H(t) \leq H(0) +  \|d(\cdot)\|_2^2\frac{1}{c_1} N^2$ with \eqref{eq_H}.
		\item (SPAV): $H(t) \leq H(0) +  \|d(\cdot)\|_2^2\frac{1}{c_2} N$ with \eqref{eq_H}.
		\item (APAV): $H_\Delta(t) \leq H_\Delta(0) +  \|d(\cdot)\|_2^2\frac{1}{c_3} c^N$ with \eqref{eq_Hdelta}.\\This holds for $\hp \geq \hd$ and $\hd < 1$.
	\end{itemize}
	where $c_1>0, c_2>0, c_3>0$ and $c>1$ are some constants independent of $N$. For $\hd > \hp$ we conjecture instability for a sufficiently large string length $N$.
	\label{thm:scalingAll}
\end{thm} 

\begin{pf}
	{
	The scaling of the SPSV and SPAV follows by the use of \eqref{eq:scalingSym} and \eqref{eq:scalingAsym}, respectively, in \eqref{eq:boundScSPSV}. Scaling of APAV directly follows from Theorem \ref{thm:expScaling}. The conjecture about instability (Conjecture \ref{con:instab}) is based on our results in Lemma \ref{lem:instability} and numerical simulations.} $\hfill\strut$\qed
	\end{pf}
	
{Note that the bounds on $H(t)$ and $H_\Delta(t)$ above yield bounds on the state vector at a given time $t$ since the Hamiltonians are positive definite functions. Hence, results on how $H(t)$ and $H_\Delta(t)$ scale as the string length $N$ grows, directly implies how the norm of the state vectors scales with $N$.}

\subsection{Discussion of Theorem~\ref{thm:scalingAll}}
	The results of Theorem~\ref{thm:scalingAll} are illustrated in Fig. \ref{fig:completePicture}. The region denoted as $A$ corresponds to the case with symmetric coupling in position and asymmetric in velocity (SPAV). For this, it was shown that the scaling is linear in $N$. When the coupling becomes symmetric also in velocity (SPSV), i.e., point $B$, the scaling deteriorates to $N^2$. In region $C$, where the asymmetry in position is less than the asymmetry in velocity (APAV), the scaling is exponential in the worst case (Theorem \ref{thm:expScaling}). We conjecture, based on numerical simulations, that the same exponential scaling occurs also in the region $D$. In the region $E$, there exist combinations of $\hd$ and $\hp$ for which even trivial strings of length $N=2$ are unstable. We conjecture that for all $\hd>\hp$ there exists a critical stable string length beyond which the string becomes unstable.

Scaling in some of the regions were known previously. For instance, the case $\hp=\hd=1$ corresponds to the predecessor following (PF) case, for which Seiler et al.\ in \cite{Seiler2004a} proved that the $\hinfnorm$ norm grows exponentially. Later, this was generalised in \cite{Tangerman2012, Herman2013b} to $0\leq\hp=\hd\leq1$. However, these popular choices are clearly outperformed by case A, that is, choosing $\hp>0$ and $\hd=0$. This effect is also illustrated by several numerical simulations discussed above. Therefore, we believe that the results presented in this paper should lead to a new ``standard'',  that is, choosing $\hp>0$ and $\hd=0$.

\subsection{Difference between velocity and position coupling}
{Symmetric velocity coupling can be interpreted as virtual dampers, whereas symmetric position couping can be seen as virtual springs.} The dampers are instances of generalised resistances \cite{Jeltsema2009}. Hence, they only extract (``burn'') energy from the system. When introducing asymmetric dampers, only \emph{how} the energy is extracted is changed. Allowing asymmetric position coupling has a different effect: Assuming ideal springs, a force acting on one side of the spring is exactly the opposite of a force at the other side of the spring by Newton's third law (`actio = reactio'). This fundamental law is violated when introducing asymmetric position coupling: Consider $A=I$, $\Delta_i>0$ and $\hd>0$. Then, the force, which is pulling the preceding vehicle backwards, is $(1-\hd)\Delta_i$, while the force, which is pulling the following vehicle forward, is $(1+\hd)\Delta_i$. Combining them yields $2\hd\Delta_i>0$. Thus, asymmetric position coupling introduces additional forces, and hence adds energy to the system.

%
%
\subsection{Future directions}
We conjecture that claims similar to Theorem \ref{thm:scalingAll} can be made for more realistic vehicle models and maybe also for more general graph topologies.{Also the scaling of the standard system norms, $\mathcal{H}_2$ and $\mathcal{H}_\infty$, is worth investigating}.

\appendix
 
\section{Proof of Lemma~\ref{lem:scalingLowerBound}}\label{app:lemLowerBound}
	First, consider the symmetric case such that $\left(\BB+\BBpt\right) R \BBT=\BB R \BBT$. {Let $D=R^{1/2}$. Then $\BB R \BBT = (\BB D)(\BB D)^T$ and  $\smin(\BB D) \geq \sqrt{\underline r}\smin(\BB)$ \cite[Prop. 9.6.4]{Bernstein}. Then the minimal singular value (= minimal eigenvalue) can be bounded by 
	\begin{equation}
		\lmin\left(\BB R \BBT\right) \geq \underline r \lmin\left(\BB \BBT\right), \label{eq:Rexclude}
	\end{equation}}
with $\BB \BBT$ being a pinned Laplacian for an undirected path graph. Its eigenvalues are given as \cite[Prop. 3.3]{Parlangeli2012}	
%
		%
	%
	\begin{equation}
		\lambda_i = 2\left(1-\cos \frac{(2i-1) \pi}{2 N + 1} \right) = 4 \sin^2 \frac{(2i-1) \pi}{4N+2}, \:\: i=1, \ldots, N.
		\end{equation}
	{The smallest eigenvalue $\lambda_{1}$ is bounded using $\sin x \geq 2x/\pi$ as
	\begin{equation}
		\lambda_1(\BB \BBT) = 4 \sin^2 \frac{\pi}{4N+2} \geq  4 \left( \frac{1}{2N+1} \right)^2 \geq \frac{1}{4}\frac{1}{N^2}. 
		\label{eq:lminBBBound}
		\end{equation}
	Then we get the quadratic bound
	\begin{equation}\label{eq:lminBBBound2}
		\lmin \BB R \BBT \geq \underline r \frac{1}{4}\frac{1}{N^2}.
	\end{equation}}
		%
%
	
	To determine the decay of the smallest singular value for the case $\hp>0$, denote $ \lp = (\BB + \hp\langle\BB\rangle)  R \BBT$. Then
	\begin{align}
		\sigma_\mathrm{min}^2 & \left((\BB + \hp\langle\BB\rangle)  R \BBT\right) = \lmin \left(\lp^\T \lp\right)
		\end{align}
	The smallest eigenvalue of the matrix can be 
	rewritten as 
	\begin{align}
		\nonumber\lmin \left(\lp^\T \lp\right) =& \lmin \Big[\left(\BB R \BBT\right)^\T\left(\BB R \BBT\right) 
		 	+ \hp^2 \BB R \langle\BB\rangle^\T\langle\BB\rangle R \BBT \\
		\nonumber & +\hp \left( \BB R \left(\langle\BB\rangle^\T\BB  + \BBT \langle\BB\rangle\right) R \BBT \right)\Big]\\
		\geq& \underline r^2  \lmin \left(\Gamma_1 + \hp \Gamma_2 + \hp^2 \Gamma_3 \right). 
		\end{align}
	with $\Gamma_1 = (\BB \BBT)^\T(\BB \BBT)$, $\Gamma_2 = \left( \BB \left(\langle\BB\rangle^\T\BB  + \BBT \langle\BB\rangle\right) \BBT \right) = \mathrm{diag}(2, 0, .., 0)$ and $\Gamma_3 = \BB \langle\BB\rangle^\T\langle\BB\rangle \BBT$. Restructuring yields
	\begin{equation}
 		\lmin \left(\lp^\T \lp\right) 
		\geq \underline r^2 \lmin \left(\Gamma_1 + \hp^2 \Psi_1 + \hp^2 \Psi_2 \right),
		\end{equation}
	where
	\begingroup
\begin{align}
		\Psi_1 &= \begin{bmatrix}
				1+\frac{2}{\hp} & 0 & -1 & 0 & 0  & \ldots & 0 \\
				0 & 2 & 0 & -1 & 0 & \ldots & 0 \\
				-1 & 0 & 2 & 0 & -1 & \ddots & \vdots\\
				0 & \ddots & \ddots & \ddots & \ddots & \ddots & 0\\
				\vdots & \ddots & -1 & 0 & 2 & 0 & -1 \\
				0 & \ldots & 0 & -1 & 0 & 1 & 0 \\
				0 & \ldots & 0 & 0 & -1 & 0 & 1 \\
			\end{bmatrix}
	\end{align}
\endgroup
	{and $\Psi_2$ is a matrix of zeros with $\begin{bmatrix} 1 & -1 \\ -1 & 1\end{bmatrix}$ in the bottom-right corner}. Using \cite[Fact 5.12.2]{Bernstein}, $\lmin$ can be bounded by
	\begin{equation}
		\lmin \left(\lp^\T \lp\right) \geq \underline r^2\left(\lmin \Gamma_1 + \hp^2 \lmin (\Psi_1) + \hp^2 \lmin (\Psi_2)\right). \label{eq:eigLowBound}
		\end{equation}
	By \eqref{eq:lminBBBound}, $\lmin(\Gamma_1)=\lmin(\BB \BBT)^2 \geq 1/(16N^4)$. The matrix $\Psi_2$ is positive semi-definite matrix, hence $\lmin (\Psi_2)=0$. It remains to investigate $\lmin (\Psi_1)$. Note that $\Psi_1$ is a reducible matrix. Using the permutation matrix $P=[\vec e_1, \vec e_3, \ldots \vec e_{N-1}, \vec e_2, \vec e_4, \ldots \vec e_N]$, leads to
	\begin{equation}
		\lmin (\Psi_1)=\lmin \left(P^{-1} \Psi_1 P\right)=\lmin \left(\begin{bmatrix} L_1 & 0 \\ 0 & L_2 \end{bmatrix}\right),
		\end{equation}
	which is a block diagonal matrix with matrices defined as $L_1 = \bar\BB D {\bar\BB}^\T \in \mathbb{R}^{N/2\times N/2}$ and $L_2 = \bar \BB {\bar\BB}^\T \in \mathbb{R}^{N/2 \times N/2}$ with $D = \mathrm{diag}(1/\hp, 1, \ldots, 1)$ and $\bar\BB$ has the same structure as $\BB$ but half the size. It follows that $\lmin (\Psi_1)=\min\left\{\lmin (L_1), \lmin (L_2)\right\}$. Let $\gamma=\min\{1/\hp, 1\}$. The eigenvalues of individual matrices are given as $\lmin (L_1) \geq \gamma \lmin \bar\BB {\bar\BB}^\T$ and $\lmin (L_2) = \lmin \left(\bar\BB {\bar\BB}^\T\right)$. Hence, $\lmin (\Psi_1) \geq \gamma \lmin \left(\bar\BB {\bar\BB}^\T\right)$. Since $\bar \BB$ has the size $N/2 \times N/2$, from \eqref{eq:lminBBBound} we get $\lmin(\Psi_1) \geq \gamma \big/{N^2}$.
	{Using this, \eqref{eq:lminBBBound}, \eqref{eq:eigLowBound} and $\lmin(\Psi_2)=0$ yields the final result \eqref{eq:scalingLowerBound}.
}

\section{Proof of Lemma~\ref{lem:scalingSigma}}\label{app:lemScalingSigma}
	{The lower bound on the smallest singular value is given by \eqref{eq:scalingLowerBound}. For $\hp=0$, quadratic scaling is shown in \eqref{eq:lminBBBound2}. When $\hp > 0$, the approach with rate ${1}/{N^4}$ is much faster than the approach of $\hp^2 \gamma /{N^2}$, hence for $N$ sufficiently large 
		$\smin((\BB+\BBpt)R\BB) 
		\geq \underline r \hp \sqrt{\gamma}\big/{N}.$
	}
	{To obtain the upper bound, first note that 
	\begin{align}
		\nonumber &\smin\left(\left(\BB+\BBpt\right)R \BB^\T\right) = \min_x {\left\|\left(\BB+\BBpt\right)R \BB^\T x\right\|_2}\bigg/{\|x\|_2} \\ 
		\nonumber &\leq \min_x \frac{\|R\|_2\left\|\left(\BB+\BBpt\right)\BB^\T x \right\|_2}{\|x\|_2} = \smax(R)\min_x \frac{\left\|\left(\BB+\BBpt\right)\BB^\T x\right\|_2}{\|x\|_2} \\
		&= \overline r \smin\left(\left(\BB+\BBpt\right)\BB^\T\right),
	\end{align}
	where in the second step the submultiplicativity of the induced norm \cite[eq. 9.3.4]{Bernstein} was used.
	 Hence, it suffices to analyse  $\smin\left(\left(\BB+\BBpt\right)\BBT\right)$,}
	which has the form 
	\begingroup
	\begin{equation}
	\begin{bmatrix}
			2 & -(1\!-\!\hp) & 0 & \ldots & 0 \\
			-(1\!+\!\hp) & 2 & -(1\!-\!\hp) &  \ldots & 0 \\			
			\vdots & \vdots & \vdots & \ddots & \vdots \\
			 0 & \ldots &  -(1\!+\!\hp) & 2 & -(1\!-\!\hp) \\
			 0 & \ldots & 0 & -(1\!+\!\hp) & 1\!+\!\hp
		\end{bmatrix}.
		\label{eq:lapl}
	\end{equation}
	\endgroup
As its leading principal submatrix of size $N-1$, it has a finite Toeplitz matrix, denoted as $M_N$. The matrix $M_N$ has as its symbol $a(t)=-(1-\hp)t^{-1} + 2 - (1+\hp)t^{1}$ with $t\in\mathbb{C}, |t|=1$. The symbol is not Fredholm, because it has a zero at $t=1$. The order $\alpha$ of the zero at $t=1$ is either $1$ for $\hp>0$	or $2$ for $\hp=0$. The result \cite[Thm. 9.8]{Bottcher2005} specifies scaling of singular values for Toeplitz matrices as
	\begin{equation}
		\sigma_i(M_N) = \mathcal{O}\left(1/{N^\alpha}\right) \label{eq:scalingSinValsToep}
		\end{equation}
	for any fixed $i$ with $\sigma_i \leq \sigma_{i+1}$. That is, the singular values go to zero with a rate at least given by the order of the zero of the	symbol. 
	Since $M_N$ is a submartix of $\left(\BB+\BBpt\right)\BBT$, use the result \cite[Thm. 9.7]{Bottcher2005} on interlacing of the singular values for submatrices. It follows that
	\begin{equation}
		\sigma_{\min}\left(\left(\BB+\BBpt\right)\BBT\right) \leq \sigma_3 (M_N). \label{eq:sinValInterlacing}
		\end{equation}
	From \eqref{eq:scalingSinValsToep} follows that $\sigma_3(M_N)=\mathcal{O}\left({1}/{N^\alpha}\right)$, hence $\sigma_3(M_N) \leq c/N^\alpha$. Thus, by \eqref{eq:sinValInterlacing} $\sigma_{\min}\left(\left(\BB+\BBpt\right)\BBT\right) \leq c_2/N$ if $\hp>0$ and $\sigma_{\min}\left(\left(\BB+\BBpt\right)\BBT\right) \leq
	c_4/N^2$ if $\hp=0$. $\hfill\strut$\qed

\section{Proof of Theorem~\ref{thm:expScaling}}
\label{app:APAV}
First note that the system model \eqref{eq:portHamAsymPos} is a skew-symmetric form, since $-(-\BBT E^{-1})^\T=E^{-1}\BB$ is equal to $\frac{1}{1+\hd}(\BB+\BBdt)E^{-1}$. Hence, it is a proper port-Hamiltonian form. To see this, rewrite the latter as $\left(\frac{1}{1+\hd} \BB + \frac{\hd}{1+\hd} \Babs \right)E^{-1}$. It can be easily verified that $\left(\frac{1}{1+\hd} \BB + \frac{\hd}{1+\hd} \Babs \right) = I - \frac{1-\hd}{1+\hd} D_\mathrm{u}$, where $D_\mathrm{u}$ has ones only at the first upper-diagonal. Also, $\frac{1-\hd}{1+\hd} D_\mathrm{u}E^{-1} = E^{-1} D_\mathrm{u}$. This yields $\left(\frac{1}{1+\hd} \BB + \frac{\hd}{1+\hd} \Babs \right)E^{-1} = (I \!-\! \frac{1-\hd}{1+\hd} D_\mathrm{u}) E^{-1} = E^{-1} - E^{-1} D_\mathrm{u} = E^{-1}\BB$. 

	(i) Use $H_\Delta$ in \eqref{eq_Hdelta} as a Lyapunov function and set $d(t)=0$. With $\vt:= M^{-1}(p-M\underline 1v_0)$ such that $\nabla_p H_\Delta = E\vt$, the time derivative is $\dot H_\Delta = -\vt^\T E \left(\BB+\BBpt\right) R \BBT \vt$. This quadratic form is equivalent to the quadratic form $-\vt^\T S \vt$ with $S=\frac{1}{2}\left(E \left(\BB+\BBpt\right) R \BBT +  \BB R (\BB+\BBpt)^\T E\right)$, $S=S^\T$. Thus, $-\vt^\T E \left(\BB+\BBpt\right) R \BBT \vt < 0$ for all $\vt$ if and only if $S>0$. The sum $s_i$ of the $i$th row of $S$ is $s_i = \left((\hp-\hd)\big(r_i(1+\hd)-r_{i+1}(1-\hd)\big)(1-\hd)^{i-2}\right)/{(1+\hd)^i}$
and the sums $s_1 = \Big((1+\hd+\hp+\hd\hp)r_1 - r_2(\hp-\hd)\Big)/\Big(1+\hd\Big)$ and $s_N = r_N ((\hp-\hd) (1-\hd)^{N-2})/{(1+\hd)^{N-1}}$. Recall that by Assumption \ref{ass} $r_i \geq r_{i+1}$. Then if $\hp > \hd$ and $\hd < 1$, all sums $s_i$ are positive, so $S>0$. Then, $\dot H_\Delta\leq0$ and the invariance principle completes the proof. For $\hp=\hd < 1$ see \cite[Thm 2.3]{Tangerman2012}.

	(ii) Consider $d(t) \neq 0$. The derivative of $H_\Delta$ is $\dot H_\Delta = -\vt^\T E (\BB+\BBpt) R \BBT \vt + \vt^\T E d$. It can be bounded since $\dot H_\Delta \leq - \smin\left(E \left(\BB+\BBpt\right) R \BBT\right)|\vt|^2 + \vt^\T  E d$. Completing the squares then leads to a similar form as in \eqref{eq_bound_D} such that
	{
	\begin{align} 
		 \dot H_\Delta 
		\leq & \frac{|Ed|_2^2}{2\smin\left(E \left(\BB+\BBpt\right) R \BBT\right)} \leq \frac{|d|_2^2}{2\smin\left(E \left(\BB+\BBpt\right) R \BBT\right)}.
	\end{align}}
	In the last step we used the fact that $\sigma_{\max}(E)=1$. The result \eqref{eq:distBoundAPAV} follows from integrating both sides with respect to time.
	
	The smallest singular value of $E \left(\BB+\BBpt\right) R \BBT$ can be upper bounded as \cite[Prop. 9.6.6]{Bernstein} 
	\begin{align}
		\sigma_{\min}\left(E \left(\BB\!+\!\BBpt\right) R \BBT\right)\! &\leq\! \sigma_{\min}(E) \sigma_{\max}\left(\left(\BB\!+\!\BBpt\right) R \BBT\right). 
\end{align}
By Gershgorin Theorem, 
	$\smax \left(\left(\BB+\BBpt\right) R \BBT\right)\leq \bar r \sigma_{\max}((\BB+\BBpt)\BBT) \leq 4 \bar r$.
Also, $\smin(E)=\left(\frac{1-\hd}{1+\hd}\right)^{N-1}$. Then, 
\begin{align}
		\smin \left(E \left(\BB+\BBpt\right) R \BBT \right)	&\leq \left(\frac{1-\hd}{1+\hd}\right)^{N-1} 4 \bar r \propto \frac{1}{c^N}.
		\end{align}
	with $c=\frac{1+\hd}{1-\hd} > 1$.  Thus, $\smin \left(E \left(\BB+\BBpt\right) R \BBT \right)\bigg)$ goes to zero exponentially fast. $\hfill\strut$\qed

\small
\bibliographystyle{plain}
\bibliography{Papers-AsymmetryPortHamApproach}

\end{document}